\documentclass[12pt,twocolumn]{iopart}
\usepackage{graphicx}
\usepackage{bm}
\usepackage{color}
\begin{document}
\title{Ising $t$--$J$ model close to half filling: A Monte Carlo study}
\author{M M Ma{\'s}ka$^1$, M Mierzejewski$^1$, A Ferraz$^2$, E A Kochetov$^3$}
\address{$^1$ Department of Theoretical Physics, Institute of Physics, University of
Silesia, 40--007 Katowice, Poland}
\address{$^2$ International Center of Condensed Matter Physics,
Universidade de Brasilia, Caixa Postal 04667, 70910--900 Brasilia,
DF, Brazil}
\address{$^3$ Laboratory of Theoretical Physics, Joint Institute for
Nuclear Research, 141980 Dubna, Russia}

\begin{abstract}
Within the recently proposed doped-carrier representation of the projected
lattice electron operators we derive a full Ising version of the $t$--$J$ model. This model
possesses the global discrete ${Z}_2$ symmetry as a maximal spin symmetry
of the Hamiltonian at any values of the coupling constants, $t$ and $J$.
In contrast, in the spin anisotropic limit of the $t$--$J$ model, usually referred to as the
$t$--$J_z$ model, the global $SU(2)$ invariance is fully restored at $J_z=0$,
so that only the spin-spin interaction has in that
model the true Ising form. We discuss a relationship between those two models and the
standard isotropic $t$--$J$ model. We show that the low-energy quasiparticles in all
three models share
the qualitatively similar properties at low doping and small values of $J/t$.
The main advantage of the proposed Ising $t$--$J$ model over the  $t$--$J_z$ one is that
the former allows for the unbiased Monte Carlo calculations on large clusters of up to
$10^3$ sites. Within this model we discuss in detail
the destruction of the antiferromagnetic (AF) order by doping as well as
the interplay between the AF order and hole mobility.
We also discuss the effect of the exchange interaction and that of the next
nearest neighbour hoppings on the destruction of the AF order at finite doping.
We show that the short-range AF order is observed in a wide range
of temperatures and dopings, much beyond the boundaries of the AF phase.
We explicitly demonstrate that the local no double occupancy constraint
plays the dominant role in destroying the magnetic order at finite doping.
Finally, a role of inhomogeneities is discussed.
\end{abstract}
\maketitle

\section{Introduction}

Since the discovery of the high--temperature superconductors (HTSC),
many theoretical investigations have been focused on the study of the doping
evolution from the antiferromagnetically ordered Mott insulator to the BCS-type
superconductor.
It is commonly assumed that this evolution is the key ingredient for
the understanding of the physics behind high--temperature superconductivity
and it can adequately be addressed in terms of the two--dimensional ($2d$) $t$-$J$ lattice model.
This model is believed to capture the essential low-energy physics of
doped Mott insulators driven by strong electron correlations.
Although, it is generally accepted that strong electron
correlations play an important role close to half filling, it remains unclear to what
extent HTSC can be described entirely in terms of this minimal
electronic model.

One of the reasons for this lack of clarity is that, despite its
simplicity, the exact properties of the $t-J$ model, apart from a few
limiting cases, are still unknown. The vast majority of the results
away from half filling have been obtained for one or two holes
introduced into the AF background. This problem has
been thoroughly analyzed with the help of various analytical and
numerical approaches \cite{onehole1,onehole2,onehole3,
onehole4,wrobel,onehole5,onehole6,onehole7,onehole8,onehole9,bonca1}.
Results for larger dopings are less comprehensive. Here, one of the
important problems concerns the robustness of the AF
order against doping. Most of the theoretical approaches predict,
that the long range AF order persists up to much
larger dopings than  observed in cuprates \cite{lugas,morinari}. It
has recently been suggested that including into consideration
the nearest neighbour hopping may help in resolving this discrepancy \cite{spanu}.
However, a reliable and well controlled analytical treatment of the isotropic
$t$--$J$ model poses a severe technical problem: it is very hard to analytically
deal with the local no double occupancy (NDO) constraint. On the other hand,
because of this constraint, numerical treatment is available
only on rather small lattice clusters, which immediately rises the problem of the finite-size
effects and, consequently, of the thermodynamical stability of the results obtained.

An interesting question then arises: is there available a modification of the isotropic $t-J$ model that
allows for unbiased numerical treatment and
at the same time captures the essential properties of the original $t-J$ model at least for a certain
parameter range? Some of the computational difficulties related with the
$t$--$J$ model may be overcome by means of investigations
of its anisotropic limit, i.e., the $t$--$J_z$ model.
Numerical work has suggested that the $t-J$ and $t-J_z$ models have
many similar properties \cite{dagrmp}.
In particular, the stripes, pairing, and Nagaoka states
found in the $t-J_z$ model are very similar to those of the isotropic
$t-J$ model \cite{cherwhite,cherwhite1,riera,martins,cz1}. In general,
despite some significant differences between both the models \cite{cz1,tjbw},
at the low-energy scale of order $J\,(\ll t)$ it is reasonable to consider the background
spin configuration to be frozen with respect to the hole dynamics time scale.
In this case the properties of the low-energy quasiparticle excitations in the $t-J$ model
are at least qualitatively similar to those in the anisotropic $t-J_z$ model.
Although this model is more amenable to numerical calculations, again only rather small lattice clusters
are allowed.

One may hope that the full Ising version of the $t-J$ model in which the $t$-term possesses the global
discrete $Z_2$ spin symmetry rather than the SU(2) one results in a more tractable though still nontrivial model.
However, it is not clear how such
a model can be derived directly in terms of the Gutzwiller projected lattice electron operators, since those operators
transform themselves in a fundamental representation of SU(2).
In this paper we use the recently proposed doped-particle representation of the projected electron
operators to derive the full Ising version of the $t-J$ model. We believe that the results of our study
capture some essential low-energy properties of the isotropic $t-J$ model
since the strongly correlated nature of the problem is preserved.

The doped-particle representation of the $t$-$J$
model is especially suited for investigations of the underdoped
regime \cite{ribeiro, cobalt, yourlast}. Some crucial points of this
approach are recalled at the beginning of the following section.
Although this is a slave--particle formulation, it differs from other
similar approaches in that the NDO constraint is
identically fulfilled at half--filling. Since the holes are the only
charge degrees of freedom present in the system, the number of
charge carriers is very small close to half-filling. Therefore, this
approach is particularly useful for the description of the
underdoped regime, whereas its applicability to the description of
strongly overdoped cuprates is much more involved.

Until now, the doped particle representation  has been
analyzed only within mean--field  approximations \cite{ribeiro,
cobalt}. Our aim is to go beyond this approximation in such a way,
that the slave--particle constraint is exactly taken into account.
We demonstrate that this can be achieved for
very large clusters of the order of $10^3$ lattice sites,  provided
the SU(2) symmetry is broken down to $Z_2$ not only in the spin--spin interaction term
but also in the hopping term.  Since the
effectiveness of the  resulting calculations is independent of the
translational invariance, this approach also allows one to
investigate the role of inhomogeneities which are expected to play
an important role in the cuprate compounds \cite{dagotto}.

The paper is organized as follows. Section 2 describes in detail
the derivation of the theoretical model. Section 3 comprises the
results of the numerical calculations. Section 4 lists
our conclusions.

\section{Model and Approach}

\subsection{Doped-carrier formulation of the $t$--$J$ model}

We start with the $t$--$J$ Hamiltonian on a square lattice \cite{spalek}
\begin{equation}
H_{tJ}=-\sum_{ij\sigma} t_{ij} \tilde{c}_{i\sigma}^{\dagger}
\tilde{c}_{j\sigma}+ J\sum_{\langle ij\rangle} (\bm Q_i \bm Q_j -
\frac{1}{4}\tilde{n}_i\tilde{n}_j),\label{2.1}
\end{equation}
where
$\tilde{c}_{i\sigma}=Pc_{i\sigma}P=c_{i\sigma}(1-n_{i,-\sigma})$ is
the projected electron operator (to exclude the on-site double
occupancy), $\bm
Q_i=\sum_{\sigma,\sigma'}\tilde{c}_{i\sigma}^{\dagger}\bm\tau_{\sigma\sigma'}\tilde{c}_{i\sigma'},
\,\bm\tau^2=3/4, $ is the electron spin operator and $\tilde
n_i=Pn_iP=n_{i\uparrow}+n_{i\downarrow}-2n_{i\uparrow}n_{i\downarrow}$.
Hamiltonian~(\ref{2.1}) contains a kinetic term with the hopping
integrals $t_{ij}$  and a potential $J$ describing the strength of
the nearest neighbour spin exchange interaction. At every lattice
site the Gutzwiller projection operator
$P=\prod_i(1-n_{i\sigma}n_{i-\sigma})$ projects out the doubly
occupied states $|\uparrow\downarrow\rangle$ thereby reducing the
quantum Hilbert space to a  product of $3$-dimensional spaces
spanned by the states $|0\rangle_i, \, |\uparrow\rangle_i$
and $|\downarrow\rangle_i.$ Physically this modification of the
original Hilbert space results in strong electron correlation
effects. The crucial local no double occupancy constraint is
rigorously incorporated into Eq. (\ref{2.1}). However, this is
achieved at the expense of introducing the constrained electron
operators, $\tilde{c}^{\dagger}_{i\sigma}$, that obey much more
complicated commutation relations than the conventional
"unconstrained" fermion operators. It should be stressed that it is
precisely close to half filling where the Gutzwiller projection is
of a crucial importance: the projected electron operator $\tilde
c^{\dagger}_{i\sigma}$ in this regime significantly differs from the
bare electron operator $c^{\dagger}_{i\sigma}$ (right at half
filling $\tilde c^{\dagger}_{i\sigma}=0$).

A natural question then arises: is it possible to rewrite the
$t$-$J$ Hamiltonian in terms of the conventional fermion and spin
operators in such a way that the NDO constraint for the lattice
electrons transforms itself into the one that can be treated, close
to half filling, in a controlled way? Recently, it has been shown
that the $t$--$J$ Hamiltonian can indeed be represented in that
form \cite{cobalt,ribeiro,yourlast}.

For the reader's convenience we sketch below the main points of this
scheme. The basic idea behind this approach is to assign fermion
operators to doped carriers (holes, for example) rather than to the
lattice electrons. The $t$-$J$ Hamiltonian is expressed then in
terms of the lattice spin operators, $\bm{S}_i$, and doped carrier
operators represented by spin-$1/2$ charged fermions, $d_{i
\sigma}$.

To accommodate these new operators one obviously needs to enlarge
the original onsite Hilbert space of quantum states. This enlarged
space is characterized by the state vectors $|\sigma a\rangle$ with
$\sigma=\Uparrow,\Downarrow$ labeling the spin projection of the
lattice spins and $a=0,\uparrow,\downarrow$ labeling the dopon
states (double occupancy is not allowed). In this way the enlarged
Hilbert space becomes
\begin{equation}
{\cal H}^{enl}_i=\{|\Uparrow 0\rangle_i,|\Downarrow
0\rangle_i,|\Uparrow \downarrow\rangle_i, |\Downarrow
\uparrow\rangle_i,|\Uparrow \uparrow\rangle_i,|\Downarrow
\downarrow\rangle_i\},
\end{equation}
while in the original Hilbert space we can either have one electron
with spin $\sigma$ or a vacancy:
\begin{equation} {\cal H} =\{|\uparrow \rangle_i,|\downarrow
\rangle_i,|0\rangle_i\}\label{3v}.
\end{equation}
The following mapping between the two spaces is then defined \cite{ribeiro}:
\begin{equation}
|\uparrow \rangle_i \leftrightarrow |\Uparrow 0\rangle_i, \quad
|\downarrow \rangle_i \leftrightarrow |\Downarrow 0\rangle_i,
\end{equation}
\begin{equation} |0 \rangle_i \leftrightarrow \frac{|\Uparrow \downarrow\rangle_i
- |\Downarrow \uparrow\rangle_i}{\sqrt{2}}\label{vacancy}.
\end{equation}
The remaining states in the enlarged Hilbert space, $\left(|\Uparrow
\downarrow\rangle_i + |\Downarrow
\uparrow\rangle_i\right)/\sqrt{2}$, $|\Uparrow \uparrow\rangle_i$,
$|\Downarrow \downarrow\rangle_i$ are unphysical and should
therefore be removed in actual calculations. In this mapping, a
vacancy in the electronic system corresponds to a singlet pair of a
lattice spin and a dopon, whereas the presence of an electron is
related to the absence of a dopon.

The spin--dopon representation of the $t$-$J$ Hamiltonian then
reads \cite{yourlast}
\begin{eqnarray}
H_{t-J}&=& \sum_{ij\sigma} 2t_{ij}
\tilde{d}_{i\sigma}^{\dagger} \tilde{d}_{j\sigma}
+ J\sum_{\langle ij\rangle }\left[\left(\bm S_i + \bm M_i\right)\left(\bm S_j
+ \bm M_j\right)\right. \nonumber \\
&&-\left.\frac{1}{4}\left(1-\tilde{n}^d_i\right)\left(1-\tilde{n}^d_j\right)\right],
\label{tj1}\end{eqnarray}

\noindent with
$\tilde{d}_{i\sigma}=d_{i\sigma}(1-d_{i,-\sigma}^{\dagger}d_{i,-\sigma})$
being a projected dopon operator and
$\tilde{n}^d_i=\sum\limits_{\sigma}\tilde{d}_{i\sigma}^{\dagger}
\tilde{d}_{i\sigma}$.

The application of $H_{t-J}$ in this form should be accompanied by
the implementation of the constraint to eliminate the unphysical
states,
\begin{eqnarray}
\bm S_i\bm{M}_i +\frac{3}{4}\tilde n^d_i =0, \label{2.3}
\end{eqnarray}
where $\bm M_i=\sum_{\sigma,\sigma'}\tilde{d}_{i\sigma}^{\dagger}
\bm\tau_{\sigma\sigma'}\tilde{d}_{i\sigma'}$ stands for the dopon
spin operator so that $\bm{Q}_i=\bm{S}_i+\bm{M}_i.$ Note the
important factor of $2$ in front of the first term in Eq.
(\ref{tj1}). It originates  from the fact that the vacancies are
represented in this theory by the spin-dopon singlets given by Eq.
(\ref{vacancy}). The projected lattice electron operators can be
explicitly expressed in terms of the projected dopon operators. For
example,
\begin{equation}  \tilde{c}_{i\uparrow}^{\dagger}
=
\sqrt{2}
{\mathcal{P}}^{ph}_i  \tilde{d}_{i\downarrow}
{\mathcal{P}}^{ph}_i=
\frac{1}{\sqrt{2}} \left[\left(\frac{1}{2}+S^z_i\right)
\tilde{d}_{i\downarrow} -S_i^+ \tilde{d}_{i\uparrow} \right],
\label{eq:7}
\end{equation}
where ${\mathcal{P}}^{ph}_i=1-(\bm S_i\bm{M}_i +\frac{3}{4}\tilde
n^d_i)$ is the projection operator which eliminates the unphysical
states from the $i^{\rm th}$ site.

The above representation of the $t$--$J$ Hamiltonian is particularly
useful for the description of strongly underdoped cuprates. Close to
half filling $n^d_i \ll 1$, so that one can safely drop the tilde sign off
the projected dopon operators. This is due to the fact that in the
low doping regime the probability for the realization of a doubly
occupied dopon state is indeed very low. Despite that, the NDO
constraint (\ref{2.3}) must
be imposed to eliminate the unphysical degrees of freedom that are
present in this formalism at any finite doping. Note, however, that
at half--filling the
left hand side of Eq. (\ref{2.3}) vanishes, and thus, in contrast to
the original NDO constraint for the lattice electrons, this equation turns into a
trivial identity.\footnote{The original local NDO constraint for the lattice electrons,
$\sum_{\sigma}c^{\dagger}_{i\sigma}c_{i\sigma}\le 1,$ right at half filling reads
$\sum_{\sigma}c^{\dagger}_{i\sigma}c_{i\sigma}=1$.}
Additionally, in this regime one can neglect both the hole--hole
interactions represented by the $\bm{M}_i \bm{M}_j$ term, and the
$\tilde{n}^d_i \tilde{n}^d_j$ couplings.
Note also that the insulating phase in this representation
is directly associated with the absence of charged particles.

\subsection{The $t$--$J_z$ and the Ising $ t-J$ models}

So far, the doped--particle representation of the $t$-$J$ model has
been analyzed only within the mean--field
approximations \cite{cobalt,ribeiro,yourlast}. Our aim
is to go beyond the mean--field analysis and to
carry out calculations for large enough systems, to make sure that
the finite--size effects are truly negligible.
Let us start with the anisotropic limiting case of the $t$--$J$
model with the spins polarized only along the z-component, namely,
the $t$--$J_z$ model. This model is of interest in itself since it
captures some essential physics of strong electron correlations.

The $t$--$J_z$ model can be considered as a limiting case
of the $t$--$J$ model (\ref{2.1}) which has an Ising rather than a Heisenberg spin interaction:
\begin{equation}
H_{t-J_z}=-\sum_{ij\sigma} t_{ij} \tilde{c}_{i\sigma}^{\dagger}
\tilde{c}_{j\sigma}+ J_z\sum_{\langle ij\rangle} \left( Q^z_i Q^z_j -
\frac{1}{4}\tilde{n}_i\tilde{n}_j\right),\label{z1}
\end{equation}
Here $\tilde{c}_{i\sigma}$ represent the Gutzwiller-projected electron operators.
The global continuous spin
SU(2) symmetry of the $t$--$J$ model now reduces to the global
discrete $Z_2$ symmetry of the $t$--$J_z$ model.
Although $Q^z_i Q^z_j$ interaction possesses discrete $Z_2$ symmetry, the
original SU(2) symmetry of all other terms of the Hamiltonian is preserved.
Therefore, the symmetry of the $t$--$J_z$ model depends
on whether  $J_z$ is zero or finite. Namely, for $J_z =0$ the SU(2) symmetry
is restored again. In contrast, in the full $t-J$ model both the $t$-
and $J$-terms possess the same SU(2) symmetry.

It is therefore natural to seek for a representation of the full Ising version of the $t-J$ model
in which the symmetry of the model does not depend on the values of the model parameters.
Such a representation can straightforwardly be derived within the
dopant--particle formulation of the $t$--$J$ model.
The physical consequences as well as
computational advantages of such an approach will be discussed in the subsequent
sections.

To proceed,
we start right from the original $t$--$J$
Hamiltonian described in terms of the lattice electrons given by Eq.
(\ref{2.1}), where we now put $Q_i^{+}=Q_i^- =0$ at the operator level. We then have for the physical
electron projected operators:
\begin{eqnarray}
\tilde
c_{i \uparrow} &=& {\cal P}^{ph}_i \tilde d_{i \downarrow}^{\dagger}{\cal P}^{ph}_i
=(\frac{1}{2}+S^z_i)\tilde d_{i\downarrow}^{\dagger}, \nonumber \\
\tilde c_{i\downarrow}&=&{\cal P}^{ph}_i\tilde d_{i\uparrow}^{\dagger}{\cal P}^{ph}_i
=(\frac{1}{2}-S^z_i)\tilde d_{i\uparrow}^{\dagger},\label{1}
\end{eqnarray}
where the projection operator now reads
${\cal P}^{ph}_i=1-(2S^z_iM^z_i+\tilde n^d_i/2)$.
It can easily be checked that
$$ Q^{+}_i=(Q^{-}_i)^{\dagger} =\tilde c^{\dagger}_{i\uparrow}\tilde c_{i\downarrow}\equiv 0.$$
The kinetic $t$-term built out of the physical electron operators given by Eqs. (\ref{1})
possesses the global $Z_2$ symmetry rather than the SU(2) one.

Accordingly, the underlying onsite Hilbert space rearranges itself in the following way.
The operators $\tilde c_{i\downarrow},\,\tilde
c_{i\downarrow}^{\dagger}$ act on the Hilbert space ${\cal
H}_{\downarrow}=\left\{|\Downarrow,0\rangle,\,|\Downarrow,\uparrow\rangle\right\}$.
These operators do not mix up any other states. Operator $\tilde
c_{i\downarrow}$ destroys the spin-down electron and creates a
vacancy. This vacancy is described by the state
$|\Downarrow,\uparrow\rangle$. The similar  consideration holds for
the $\tilde c_{i\uparrow}$ operators. Now, however, the vacancy is
described by the state $|\Uparrow,\downarrow\rangle$. Those two vacancy
states are related by the $Z_2$ transformation.
The operator
$(Q^z_i)^2=\frac{1}{4}(1-\tilde n^d_i)$ produces zero upon acting on the both.
The physical Hilbert state is therefore a direct sum ${\cal
H}_{ph}={\cal H}_{\uparrow}\oplus {\cal H}_{\downarrow}.$  Under the
$Z_2$ transformation $(\uparrow\leftrightarrow\downarrow,\, S^z_i\to
-S^z_i)$ we get ${\cal H}_{\uparrow}\leftrightarrow{\cal
H}_{\downarrow}$, which results in ${\cal H}_{ph}\to {\cal H}_{ph}$.
\footnote{In the isotropic $t$--$J$ model these two $2d$ spaces
merge into a $3d$ SU(2) invariant physical space, where the vacancy is
just an antisymmetric linear combination given by the
SU(2) spin singlet~(5). The symmetric combination splits off,
since it represents an unphysical spin-triplet state.}

As a result, one arrives at the
representation (\ref{z1}) in which, however, the electron projection
operators are given by Eqs. (\ref{1}). All the parts of this Hamiltonian
possess the global discrete $Z_2$ symmetry
whereas the global continuous SU(2) symmetry is completely lost.
Close to half--filling this Hamiltonian reduces to the form,
\begin{eqnarray}
H^{Ising}_{t-J}&\equiv& H_{t-J|_{Z_2}} = \sum_{ij\sigma}t_{ij}
\tilde{d}_{i\sigma}^{\dagger} \tilde{d}_{j\sigma}
+  J\sum_{\langle ij\rangle}\left[\left(S^z_i S^z_j-\frac{1}{4}\right)  \right.
\nonumber \\
 && \left. +  S_i^zM^z_j +S_j^zM^z_i \right],
\label{tj2}
\end{eqnarray}
which should be accompanied by the constraint,
\begin{eqnarray}
2S_i^z{M}^z_i +\frac{1}{2}\tilde n^d_i =0. \label{2.31}
\end{eqnarray}
In order to emphasize the difference between the
$t$--$J_z$ and the present
model we dub the latter the Ising $t-J$ or for short the $t-J|_{Z_2}$ model.

The factor of $2$ which is presented in the hopping term of
the isotropic $t$--$J$ Hamiltonian (\ref{tj1}) drops out
from the hopping term in the Ising $t-J$ Hamiltonian.  This occurs because of the fact that
the equations
$$\tilde{c}_{i\uparrow}^{\dagger}
=\sqrt{2} {\mathcal{P}}^{ph}_i  \tilde{d}_{i\downarrow}
{\mathcal{P}}^{ph}_i, \quad  {\mathcal{P}}^{ph}_i= 1-\left(\bm
S_i\bm{M}_i +\frac{3}{4}\tilde n^d_i\right) $$
which are valid for
both the $t$--$J$ and $t$--$J_z$ models, are in the $t-J|_{Z_2}$ model replaced by the
following ones,
$$\tilde c_{i\uparrow}^{\dagger}= {\cal P}^{ph}_i\tilde
d_{i\downarrow} {\cal P}^{ph}_i, \quad {\cal P}^{ph}_i=
1-\left(2S^z_iM^z_i+\frac{1}{2} \tilde n_i^d\right).$$

In practical calculations, the NDO constraint (\ref{2.31})
can be taken into account with the help of a Lagrange multiplier.
In order to do that we
introduce an additional term to the Hamiltonian,
\begin{eqnarray}
&&\lambda\sum_i\left(2S_i^z{M}^z_i +\frac{1}{2}n^d_i\right)=  \nonumber \\
&&\lambda \sum_{i} \left[\left(\frac{1}{2}+S^z_i\right)
d^{\dagger}_{i \uparrow}d_{i \uparrow}
+
\left(\frac{1}{2}-S^z_i\right)
d^{\dagger}_{i \downarrow}d_{i \downarrow}
 \right]. \label{constr}
\end{eqnarray}
Notice that the  operator $(\frac{1}{2}+S^z_i) d^{\dagger}_{i
\uparrow}d_{i \uparrow} +(\frac{1}{2}-S^z_i) d^{\dagger}_{i
\downarrow}d_{i \downarrow}$ produces eigenvalues 0 and 1, when acting on the onsite physical and
unphysical states, respectively.
Because of this, the global Lagrange multiplier $\lambda\to \infty$
enforces the NDO constraint locally. The double dopon
occupancy of an arbitrary site results in an appearance of an
unphysical state and hence enhances the total energy by $\lambda$.
Therefore, in the large-$\lambda$ limit all unphysical states are
automatically eliminated and
we can in this limit safely remove the tilde sign off the $d$ operators.
In the following section we show that
this constraint is of crucial importance for the description of
the AF order at finite doping.

\subsection{Monte Carlo calculations}

The total  Hamiltonian takes the form
\begin{equation}
H_{t-J|_{Z_2}}^{\lambda}= H_{\uparrow}+H_{\downarrow} +J \sum_{\langle ij \rangle} S^z_iS^z_j +{\rm const},
\label{jz1}
\end{equation}
with
\begin{eqnarray}
H_{\uparrow}&=&\sum_{ij}t_{ij}
d^{\dagger}_{i \uparrow}d_{j \uparrow} \nonumber \\
&+&\sum_{i}  d^{\dagger}_{i \uparrow}d_{i \uparrow}
\left[\lambda \left(\frac{1}{2}+S^z_i\right)+\frac{J}{2}
\sum_{\langle j\rangle_i}S^z_{j}\right], \label{jz2} \\
H_{\downarrow}&=&\sum_{ij}t_{ij}
d^{\dagger}_{i \downarrow}d_{j \downarrow} \nonumber \\
&+&\sum_{i}  d^{\dagger}_{i \downarrow}d_{i \downarrow}
\left[\lambda \left(\frac{1}{2}-S^z_i\right)-\frac{J}{2}
\sum_{\langle j\rangle_i}S^z_{j}\right], \label{jz3}
\end{eqnarray}
where $\langle j\rangle_i$ denotes neighbouring sites of a given site $i$. We have neglected the
hole--hole interaction  in
$H^{\lambda}_{t-J|_{Z_2}}$, which is
perfectly justified in the low doping regime. In order to verify
this approximation we have carried out additional calculations with the
hole-hole interaction
being taken into account in the mean--field approximation. The
difference is negligible and therefore we do not present them here.
Note, that the interaction strength in
$H^{\lambda}_{t-J|_{Z_2}} $ is exactly
the same as in the standard formulation of the $t$--$J$ model.
Absence of any renormalization of the model parameters originates from the fact that
the projection procedure is explicitly
built in Eq. (\ref{jz1}), provided  $\lambda \rightarrow  \infty$.
Despite its complexity, with the Monte Carlo (MC) method, one
can investigate the Hamiltonian (\ref{jz1}) for very large systems
without any approximation. Since
$[S^z_i,H^{\lambda}_{t-J|_{Z_2}}]=0$ the spin degrees of freedom can be
analyzed within the classical Metropolis algorithm. However, since
the effective Hamiltonian (\ref{jz1}) includes both fermionic as
well as classical degrees of freedom, this algorithm needs to
be modified. The procedure is as follows:
\begin{enumerate}
\item an initial configuration of $\{S^z_i\}$ is generated;
\item the Hamiltonians (\ref{jz2}, \ref{jz3})
are diagonalized and the free energy ${\cal F}$ of the fermionic
subsystem in the canonical ensemble is determined;
\item two sites with opposite spins $S^z$ are randomly chosen;
then, both the spins are flipped;
\item step (ii) is repeated, determining new value of the free energy ${\cal F'}$;
\item if ${\cal F'} < {\cal F}$
or $\exp\left[({\cal F}- {\cal F'})/kT\right] > x$, where $x$ is a random number
from the interval $[0;\: 1)$, the new $\{S^z_i\}$ configuration is accepted,
added to the ensemble and the procedure goes to step (iii), otherwise it goes directly
to step (iii).
\end{enumerate}

It is the Metropolis algorithm, but with the internal
energy in statistical weights replaced by
the free energy of the fermionic subsystem.
A detailed description of this approach can be found in Ref.
\cite{maciek}. Concurrently with the MC simulation an
iterate procedure calculating the distribution of holes is carried
out in a self--consistent way.

Most of the numerical results have been obtained for a $20\times 20$ systems with
periodic boundary conditions. However, in order to check the influence of
finite--size effects we have carried out calculations on clusters of up to
1600 lattice sites and with averaging over the boundary conditions \cite{abc}.
This problem is discussed at the end of the next section.

In order to eliminate the unphysical states in the Monte Carlo
simulations, we have taken $\lambda=100t$. Therefore, the Lagrange
multiplier is by far the largest energy scale in the system, which
guaranties the single occupancy of each lattice site. The
simulations have been carried out in the canonical ensemble, what
allows for accurate control of the doping level. We
assume an absence of the ferromagnetic order. Namely, we take
$\sum_i S^z_i = \sum_i M^z_i=0$, what is reflected in the third point of the MC procedure.

The aim of the simulations is to determine how the antiferromagnetic
order and the spectral properties are affected by doping. We start
our discussion with the doping dependence of the spin--spin
correlation function for the projected physical electron
operators
\begin{eqnarray}
g(r)&=&\frac{4}{N^2}\sum_i\sum_j\langle (S^z_i+M^z_i) (S^z_j+M^z_j) \rangle
\nonumber \\
&\times& \exp\left[i {\bm K}\cdot ({\bm R}_i-{\bm R}_j)
\right] \bar{\delta}(r-|{\bm R}_i-{\bm R}_j|),
\label{corfun}
\end{eqnarray}
where ${\bm K}=(\pi,\pi)$ and
$$
\bar{\delta}(x)=\left\{\begin{array}{ll}
1 & {\rm if}\ |x|\le 0.5a, \\
0 & {\rm otherwise},
\end{array}\right.
$$
with $a$ being the lattice constant.
$\langle \ldots \rangle $ in Eq. (\ref{corfun}) means an average over
the spin configurations  generated in MC run.
$g(r)$ allows one to distinguish between the long range
order (LRO), when it remains finite for arbitrary $r$,
quasi long range order (QLRO), when $g(r)$ decays algebraically,
and a short range order (SRO), when $g(r)$ decays exponentially.
Calculations of the spin--spin correlation function will be accompanied by
results obtained for a static spin--structure factor, defined as
\begin{equation}
S(\bm{q})=\frac{1}{N^2}\sum_{ij}e^{i \bm{q} \left(\bm{R}_i-\bm{R}_j \right)}
\langle (S^z_i+M^z_i) (S^z_j+M^z_j) \rangle .
\label{sq}
\end{equation}
The third quantity that we use in the following discussion, is the hole
spectral function given by
\begin{equation}
A({\bm k},\omega) = -\frac{1}{\pi}{\rm Im}\: G \left({\bm k},\omega+i0^+\right),
\label{adef}
\end{equation}
where
\begin{eqnarray}
&& G \left({\bm k},z\right) = \sum_i \sum_j \exp\left\{i{\bm k} \left( {\bm R}_i
- {\bm R}_j\right)\right\}  \nonumber \\
&& \times \langle
{\cal G}_{\sigma}\left({\bm R}_i, {\bm R}_j,z \right)
\left[\frac{1}{2} -s(\sigma) S^z_i\right]
\left[\frac{1}{2} - s(\sigma) S^z_j\right]
\rangle,
\label{gdef}
\end{eqnarray}
with $s(\uparrow)=1$ and $s(\downarrow)=-1$.
Here, similarly to Eq. (\ref{corfun}), $\langle \ldots \rangle$ indicates
 averaging over spin configurations   and
\begin{equation}
{\cal G}_{\sigma}\left({\bm R}_i, {\bm R}_j,z \right)=
\left\{\left[z-H_\sigma\right]^{-1}\right\}_{ij}
\end{equation}
is the real--space Green function for a given spin configuration $\{S^z_i\}$.
The presence of factors $ \frac{1}{2} -s(\sigma) S^z_i $ in Eq. (\ref{gdef}) follows from Eq. (\ref{1}).
Note that the spin--spin correlation function, the spin--structure factor
and the spectral function
are defined for physical electron operators, $\tilde{c}_i$.
\section{Numerical results}

\subsection{Homogeneous systems}
As discussed in the preceding sections the derived
representation differs from the standard $t$--$J_z$ Hamiltonian
in that the SU(2) symmetry is broken also for $J=0$. In order to
visualize the physical consequences of this difference we start with
calculations for the one--hole case. In this regime
large clusters have been analyzed numerically both for
the $t$--$J_z$ and $t$--$J$ models.

\begin{figure}[h]
\includegraphics[width=8cm]{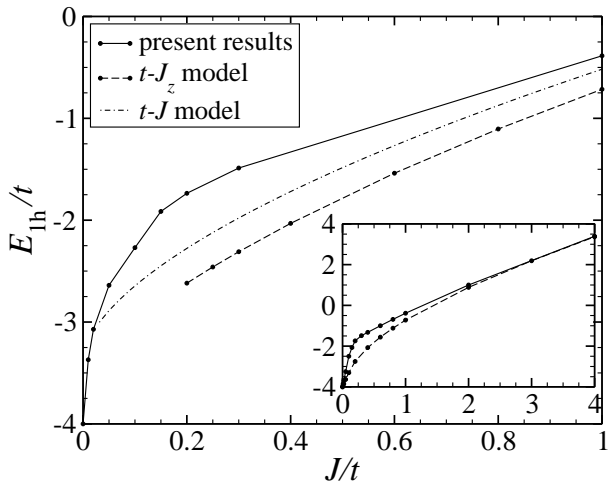}
\caption{
One--hole energy as a function of $J/t$ for a $8 \times 8$ cluster
at $T=0$. The data labeled as $t-J_z$ and $t-J$
have been taken from Refs. \cite{dagnew} and \cite{bonca1}, respectively.
In the latter case, it is the energy of a hole with momentum $(\pi/2,\pi/2)$.
The inset shows results obtained for $4\times 4$  $t-J|_{Z_2}$ and
$t$--$J_z$ systems. Here, the dashed line shows results presented in Ref.
\cite{dagnew1} for the $t$--$J_z$ model.
Since for $J \rightarrow 0$ the ferromagnetic order sets in,
the constraint $\sum_i S^z_i=0$ is now relaxed.
} \label{fig0}
\end{figure}

In the main panel of Fig. \ref{fig0}
we show the one--hole energy calculated at $T=0$ for a $8\times 8$
cluster without the
$\frac{1}{4} \tilde{n}_i \tilde{n}_j$ term. Here, we compare our data
with exact results obtained for a 50--site $t$--$J_z$
cluster \cite{dagnew} as well as with recent exact results for a bulk $t$--$J$
system \cite{bonca1}. In the inset of Fig. \ref{fig0}
we compare exact results obtained for $4\times 4$  $t-J|_{Z_2}$ and
$t$--$J_z$ clusters \cite{dagnew1} for a wider range of the $J/t$ ratio.
%
%
In two limiting cases $J/t\rightarrow 0$ and $J/t  \rightarrow \infty$
the one--hole energy obtained in our approach is the same as in the $t$--$J_z$ model.
It can be explained in the following way. For $J=0$, the ferromagnetic
Nagaoka state becomes a ground state in both the approaches. In this
case the propagation of a hole is not perturbed by the magnetic order and the
one--hole energy equals $-4t$. The main difference between $t$--$J_z$ and $t-J|_{Z_2}$ approaches
consists in the symmetry of the hopping term. Therefore, they merge
in the case $J/t \gg 1$, when the system properties are determined predominantly
by the same spin--spin interaction. In the regime of intermediate $J$ the differences
are most pronounced.
The one--hole energy
obtained for the $t$--$J$ model in this regime
is in between the results obtained for $t$--$J_z$ and $t-J|_{Z_2}$ approaches.
Here, the differences between $t$--$J$ and $t$--$J_z$ models are comparable
to those between $t$--$J$ and $t-J|_{Z_2}$ ones.

\begin{figure}[h]
\includegraphics[width=8cm]{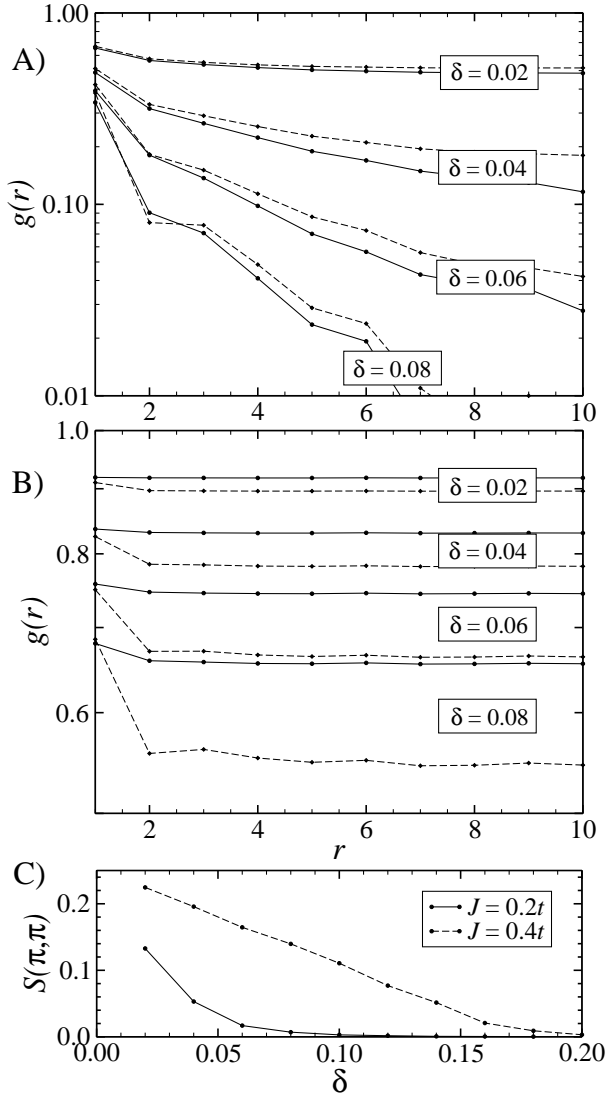}
\caption{Panels A) and B) show $g(r)$ calculated for $kT=0.1t$. $J=0.2t$ (A) and  $J=0.4t$ (B).
The curves from the top to the bottom have been obtained for
$\delta=0.02,0.04,0.06,0.08$. Solid (dashed) lines show results
obtained for $t'=t''=0$ ($t'=-0.27t$ and  $t''=0.2t$).
Panel C) shows doping dependence of the
static spin--structure factor $S(\pi,\pi)$ at $kT=0.1t$ for
$t'=-0.27t$ and  $t''=0.2t$.
} \label{fig1}
\end{figure}

Next, we investigate how the doping affects the antiferromagnetic order.
The previous studies of
the $t$--$J$ model  clearly indicate that the  long range hopping
amplitudes significantly modify the bandwidth and the dispersion of
the quasiparticles \cite{qp}. Recent Green's function Monte Carlo
calculations demonstrate that the next nearest neighbour hopping
reduces the critical doping at which the  AF LRO
disappears \cite{spanu}. The importance of these results follows from
the fact that in the $t$--$J$ model with  only the nearest neighbour
hopping, antiferromagnetic correlations persist up to hole
concentrations much larger than the ones observed in HTSC materials.
One may expect that in the absence of the transverse spin--spin
interaction the robustness of LRO should be even more pronounced.
Moreover, the previous analysis of the one-- and two--hole spectra
in the $t$--$J_z$ model \cite{cz1} has shown that for $t/J<5$, half
of the one--hole band width does not exceed $0.08t$. Therefore, the
intra--sublattice hopping should be a source of an important
contribution to the kinetic energy even for small values of $t'$ and
$t''$. The significance of the long range hopping for the AF order
is demonstrated in Fig. \ref{fig1}, where we compare the spin--spin
correlation functions calculated  with and without $t'$ and $t''$.
In all the figures showing $g(r)$ we use logarithmic scale for the
vertical axis. Therefore, for LRO, QLRO and SRO,  $g(r)$ should be
represented asymptotically by a constant function, logarithmic
function and a straight line, respectively. In the following $\delta$
denotes the average concentration of holes.

One can see that the influence of the long range hopping depends,
even qualitatively, on the value of the exchange coupling $J$. For a
small value of $J$, the AF order is {\em enhanced} when hoppings to
second and third nearest neighbours are allowed (for $J=0.2t$ see
panel A) in Fig. \ref{fig1}). On the other hand, for bigger
values of $J$, these hoppings {\em reduce} the AF order (for
$J=0.4t$ see panel B) in Fig. \ref{fig1}). Such a behaviour
could be explained as follows. In the presence of only nearest
neighbour hopping there is a strong competition between the energy of
spin--spin interaction and the hole kinetic  energy. It results from
the fact, that in this case only inter sublattice hopping is
allowed. From Eq. (\ref{constr}) one can then infer that it is
possible only in regions where the AF order is absent. Then, nonzero
$t'$ and $t''$ allow for intra sublattice hopping, thereby leading
to gaining of the kinetic energy without destroying the AF order.
This mechanism is effective for $J \le 0.25 t$. On the other hand,
for $t'=t''=0$  and large $J$, holes are almost localized and,
therefore, only weakly frustrate the AF state. The intra--sublattice
hopping allows for the propagation of holes, which effectively
reduces the AF LRO.

The doping induced destruction of the AF LRO can
directly be seen in  Fig. \ref{fig1}C, where we present
spin--structure factor obtained for $t'=-0.27t$ and  $t''=0.2t$.
This quantity is important in that it is directly accessible
in, e.g., neutron scattering experiments. The maximal doping
for which the AF state still exists strongly depends on the magnitude
of the exchange interaction. This result contrasts with the recently
reported Green's function Monte Carlo study of the $t-J$ model \cite{spanu}, where
the AF LRO vanishes at $\delta=0.1$ and $\delta=0.13$ for $J=0.2t$ and
$J=0.4t$, respectively. In our approach the experimental data for the critical
doping in HTSC can be reproduced provided $J<0.2t$.

\begin{figure}[h]
\includegraphics[width=8.5cm]{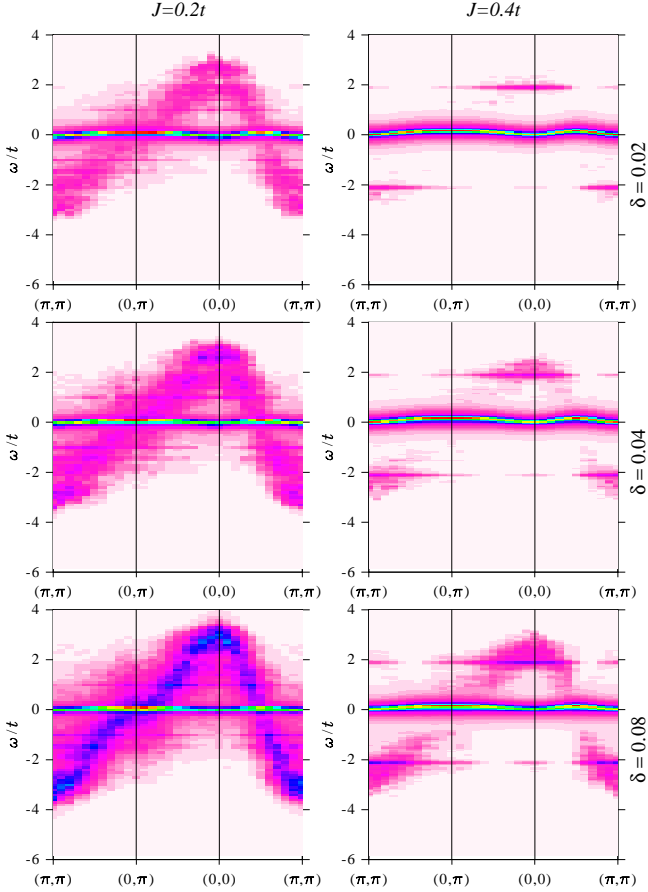}
\caption{Spectral functions $A({\bm k}, \omega)$ calculated for
$kT=0.1t$ along the main symmetry lines of the Brillouin zone with
$t'= t''=0$.} \label{fig2}
\end{figure}

\begin{figure}[h]
\includegraphics[width=8.4cm]{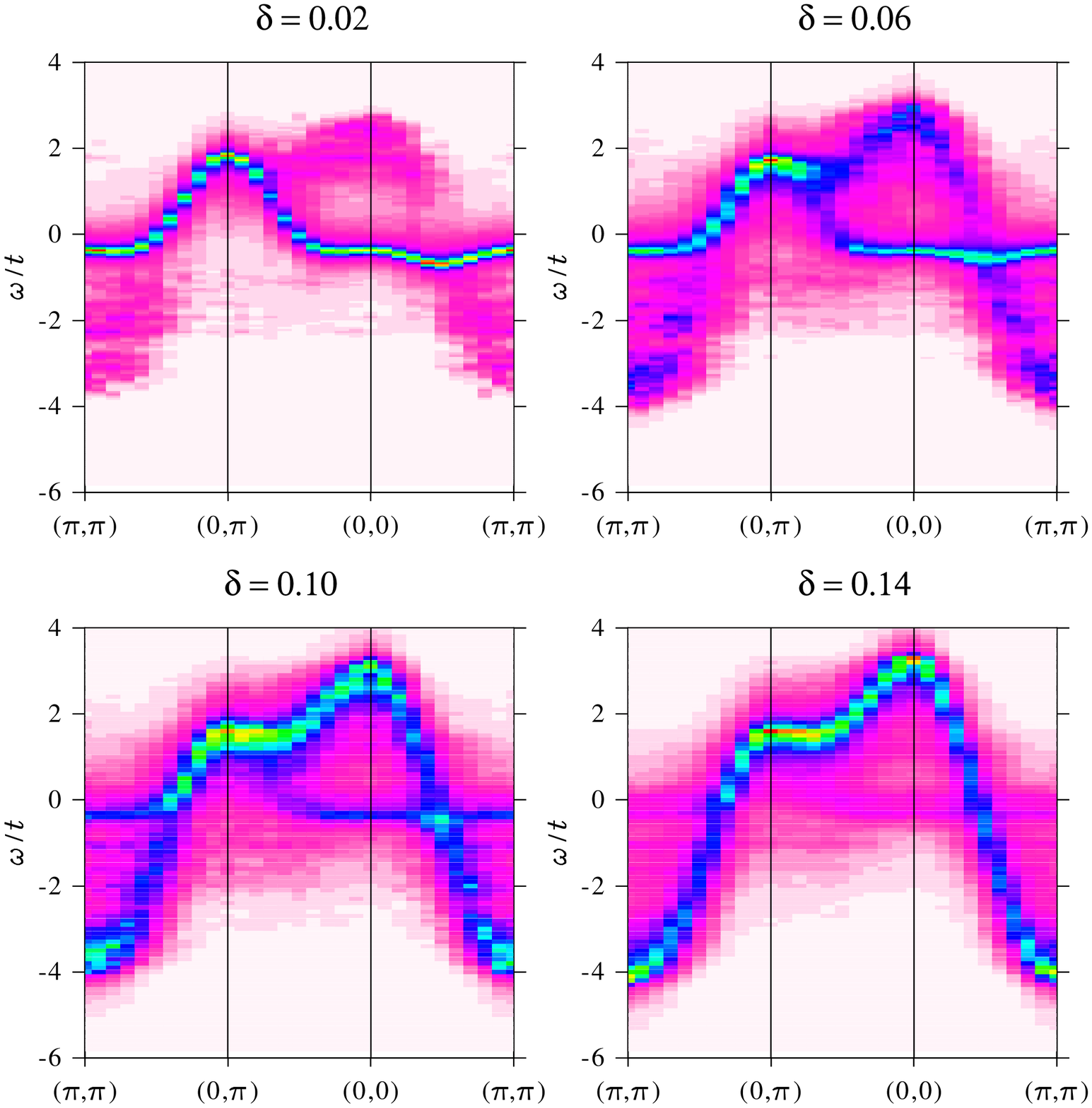}
\caption{Spectral functions $A({\bm k}, \omega)$ calculated along
the main symmetry lines of the Brillouin zone for $J=0.2t$,
$kT=0.1t$ and  various dopings.  $t'=-0.27t$ and  $t''=0.2t$ have been assumed.} \label{fig6}
\end{figure}

In order to illustrate the interplay between the AF order and
the mobility of  holes we have calculated the hole spectral functions
$A({\bm k}, \omega)$ (see Figs. \ref{fig2} and \ref{fig6}).
For  $t'=t''=0$ and small $\delta$ one can see almost localized
particles with very small dispersion. Similar situation occurs
in the $t$--$J_z$ model, but it is not the case for the $t-J$ one, where the
the spin--flip term can undo the defects generated by a moving hole
and hence it allows for its much higher mobility.
In the present approach holes become mobile
when doping increases, i.e., when the AF background disappears. The
onset of mobile holes is accomplished through a gradual transfer of
the spectral weight from the vicinity of the almost localized level.
Close to half--filling the most significant transfer takes place in
states with  ${\bm k}=(0,0)$ and ${\bm k}=(\pi,\pi)$.
However, even for relatively large doping the spectral functions remain broad
for all the momenta.

For non--zero $t'$ and $t''$ there are mobile
holes even for small doping, but the spectral functions still remain
very broad. Also in this case, doping is responsible for significant
modification of the dispersion relation of holes.
In Fig. \ref{fig6} we compare $A({\bm k},\omega)$ calculated
for $\delta=0.02,..,0.14$ with $t'=-0.27t$ and $t''=0.2t$.
Along with the destruction of the AF LRO, there is an increasing
contribution of nearest neighbour hopping to the hole kinetic
energy. For  $\delta=0.02$ the peaks in spectral functions can be
fitted by the dispersion relation with $t=0$, whereas for
$\delta=0.14$ the AF correlations hardly influence the nearest
neighbour hopping. Note that such a substantial modification of the
dispersion relation may change the topology of the Fermi surface.
Doping affects not only the effective dispersion relation, but also
frustrates the AF background. The latter effect is responsible for
strong broadening of the spectral functions that is visible in Fig.
\ref{fig6}.
Comparison of Fig. \ref{fig1} with Figs. \ref{fig2} and \ref{fig6}  demonstrates that
the mobility of holes and destruction of the AF LRO are mutually connected with each other
in the sense that mobility affects AF LRO
and, {\it vice versa}, AF order affects the mobility of holes.

\begin{figure}[h]
\includegraphics[width=8cm]{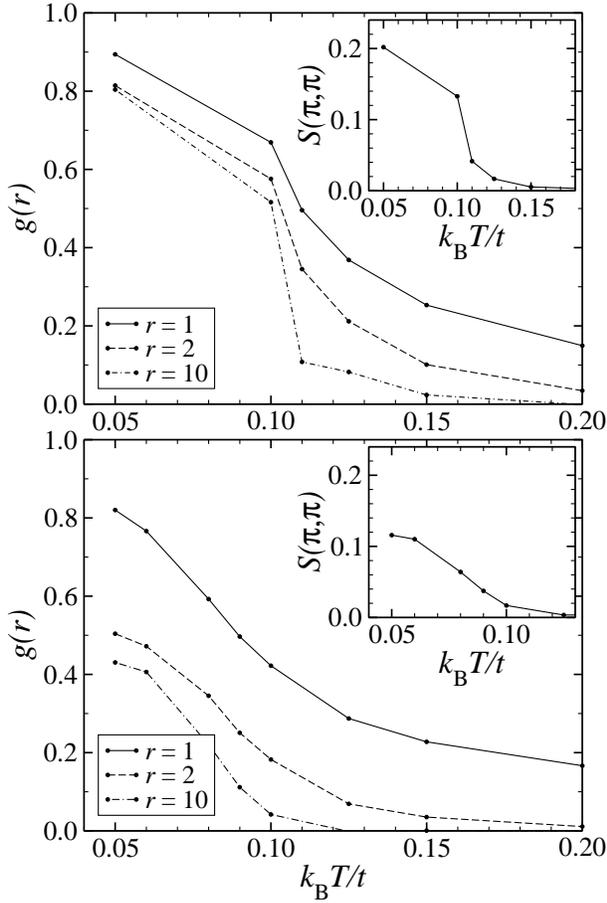}
\caption{ $g(r)$ as a function of temperature for $J=0.2t$ and
$\delta=0.02$ (upper panel) and $\delta=0.06$ (lower panel). The lines from the top
to the bottom show $g(1)$, $g(2)$ and $g(10)$. $t'=-0.27t$ and  $t''=0.2t$ have been assumed.}
\label{fig4}
\end{figure}

We now turn our attention to the temperature dependence
of the spin--spin correlation function and the spectral properties of
holes.
It is known that the N\'eel temperature drops rapidly when the parent
compounds of the high--temperature superconductors are doped with holes.
Similar behaviour can be inferred from Fig. \ref{fig4}, where
temperature
dependence of $g(r)$ is presented for different doping levels.
One can note that doping strongly reduces the LRO,
whereas its influence on the SRO is much weaker. In
particular, the nearest neighbour  correlation functions $g(1)$
calculated for $\delta=0.02$ and $\delta=0.06$ are qualitatively and
quantitatively close to each other. These results suggest that, the
AF SRO should be observed in a wide range of temperatures and
dopings, much beyond the boundaries of the AF phase.
It remains in agreement with recent experiments on high--temperature
superconductors suggesting that with doping, the long-range N\'eel order
gives way to short-range order with a progressively shorter correlation
length. As a result, at optimal doping the static spin correlation length is
no more than two or three lattice spacing \cite{exp_dop}.

\begin{figure}[h]
\includegraphics[width=8.5cm]{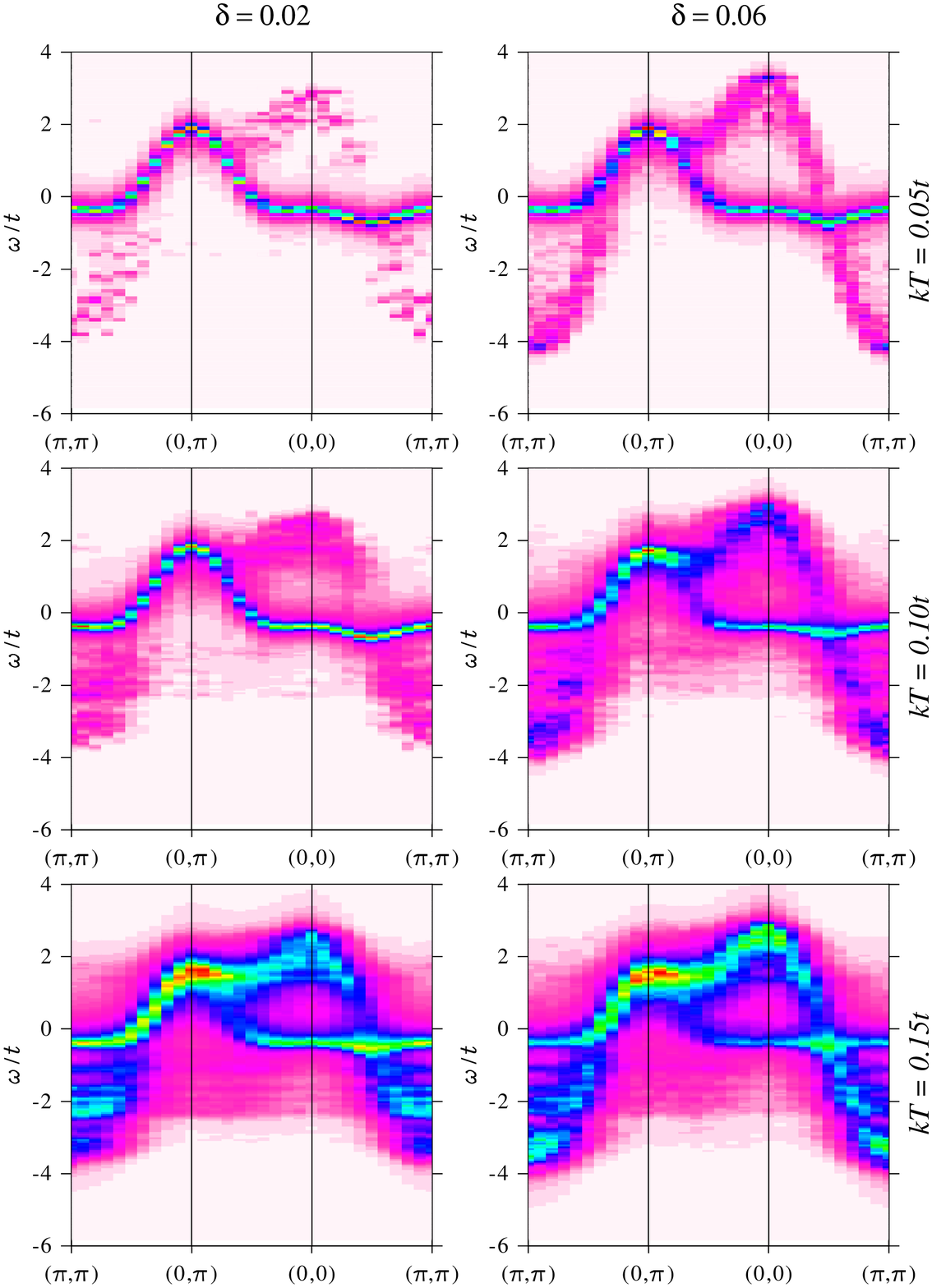}
\caption{Spectral functions $A({\bm k}, \omega)$ calculated along
the main symmetry lines of the Brillouin zone for $J=0.2t$. $t'=-0.27t$ and  $t''=0.2t$ have been assumed.}
\label{fig5}
\end{figure}

The discussed above correlation function and the spin--structure
factor describe the
background composed of the localized spins, which, as mentioned in the
preceding paragraphs, is up to some degree affected by the motion of doped holes.
Therefore, the reduction of the antiferromagnetic correlations has to be observed
also in the  dynamics of the carriers. It is shown in
Fig. \ref{fig5}, where we demonstrate the temperature dependence of the
spectral functions.
%
%
When the temperature increases, the number of
spin defects in the N\'eel state increases as well, and this enables
the nearest neighbour hopping,  thereby allowing holes to lower their
kinetic energy. This mechanism leads to trapping of holes in the regions
of broken antiferromagnetic bonds and forming ferromagnetic spin polarons,
where the hole hopping does not frustrate the spin background.
The contribution of the
nearest neighbour hopping becomes visible in the spectral functions,
where the increase of the temperature causes a significant broadening
of the spectral lines.
Similarly to the spectral functions obtained for a
single hole in the $t$--$t'$--$t''$--$J$ model \cite{brink}, the
width of the peaks in the $A({\bm k},\omega)$ is too small, when
compared to the results of the angle--resolved spectroscopy (ARPES)
measurements \cite{spec_cl} on Sr$_2$CuO$_2$Cl$_2$. It has recently
been argued, that strong electron--phonon
interaction \cite{cata,bonca2} may explain the very broad peaks
observed in the insulating copper oxides \cite{shen1,shen2}.

\begin{figure}[h]
\includegraphics[width=8cm]{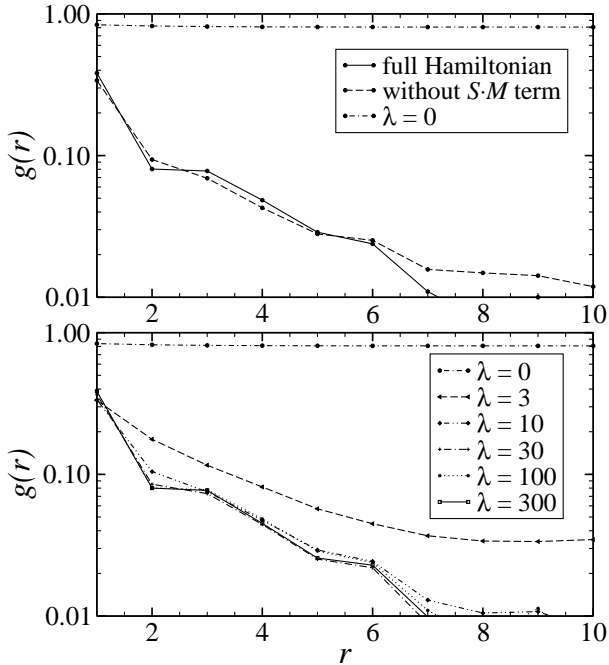}
\caption{$g(r)$ for $J=0.2t,\; kT=0.1t,\; \delta=0.08$,\; $t'=-0.27t$ and $t''=0.2t$.
In the upper panel the topmost
curve has been calculated with $\lambda=0$. The other curves show
results obtained for $\lambda=100t$ with and without the spin--hole
exchange interaction.  The lower panel shows $g(r)$ for different values
of $\lambda$. Note, that the results for $\lambda=100t$ and $\lambda=300t$
are almost indistinguishable.}
\label{fig3}
\end{figure}

In the present approach, holes
interact with spins through the intersite interaction of strength
$J$ as well as through the onsite constraint with the Lagrange
multiplier $\lambda$, and both of these interactions can be
responsible for the destruction of the AF LRO. In order to determine
the underlying mechanism we have carried out simulations taking into
account one of these interactions at a time. The results shown in
the upper panel of Fig. \ref{fig3} clearly demonstrate that the
constraint plays the dominating role in the destruction of the AF state.
When both these interactions are taken into account the LRO is
completely destroyed and only AF SRO can be observed for
$\delta=0.08$.  However, if we ignore the constraint LRO is
restored. The results obtained without the intersite
spin--hole interaction ($S_i^z M_j^z$ terms) are almost indistinguishable from the ones
obtained with both the interactions. Note, that for $\lambda=0$ the
NDO constraint is completely neglected. This illustrates the
importance of the constraint for a realistic description of the AF
order at finite doping.
Additionally, in order to determine the value of the Lagrange multiplier
for which the results converge and the NDO constraint is fulfilled we
have calculated $g(r)$ for a wide range of $\lambda$. The results are presented
in the lower panel of Fig. \ref{fig3}. One can see that the assumed value
$\lambda=100t$ is large enough to enforce the constraint.
After explaining the role of the farther neighbour hopping we restrict the
following analysis to the case of $t'=-0.27t$ and $t''=0.2t$.

\subsection{Inhomogeneous systems}

Since our method works for systems with broken translational
invariance, it is tempting to apply it as well to inhomogeneous
systems. It has recently been shown with the help of scanning
tunneling spectroscopy that nanoscale electronic inhomogeneity is an
inherent feature of many groups of high--temperature
superconductors. By a direct probing of the local density of states,
these methods reveal strong spatial modulation of the energy gap in
the superconducting Bi--based compounds \cite{Pan}. Very recently,
STM experiments have shown a strong correlation between position of
the dopant atoms and all manifestations of the nanoscale electronic
disorder \cite{McElroy,mashima}. Thus, these experiments proved
essentially that the impurities were the source of the
inhomogeneity. On the other hand, they revealed a very important
feature: there is a {\em positive} correlation between the magnitude
of a gap and the position of an out--of--plane oxygen
atoms \cite{McElroy}. These are the atoms which have been doped into
the insulating parent compound in order to introduce holes to the
CuO$_2$ planes. The gap--impurity correlation has been explained as a
result of the inhomogeneity--enhanced exchange interaction in the
$t$--$J$ model \cite{my}. Assuming that purely electronic models
contain the essential physics of cuprates, the same interaction is
responsible for both superconductivity and the AF order.  Therefore,
inhomogeneity may affect the AF state as well. Additionally,
localization of holes by the electrostatic potential of
out--of--plane oxygen atoms may also affect the AF order, since
the hopping of holes frustrates the AF LRO. In the following, we
investigate the role played by these mechanisms in the strongly
underdoped regime.
%
%
In order to investigate the latter one,
 we extend the Hamiltonian (\ref{jz1}) by adding a term responsible for
inhomogeneity--induced diagonal disorder
\begin{equation}
H^{\lambda}_{t-J|_{Z_2}} \rightarrow
H^{\lambda}_{t-J|_{Z_2}}
+\sum_{i\sigma} \varepsilon_i
d^\dagger_{i\sigma} d_{i\sigma},
\end{equation}
where
$$
\varepsilon_i=\left\{\begin{tabular}{ll}
$V$,\ \ & if there is an out--of--plane oxygen atom
\\
& above site $i$, \\
0,\ \ & otherwise.
\end{tabular}
\right.
$$

Since in HTSC each doped oxygen atom introduces one hole in the CuO$_2$ plane,
we have carried out calculations for the number of impurities equal to the number
of holes. Technically, for each Monte Carlo simulation we generate a random
configuration of the out--of--plane oxygen atoms and keep it frozen during
the whole run. In that way both holes and localized spins feel a quenched
disorder.

\begin{figure}[h]
\includegraphics[width=8cm]{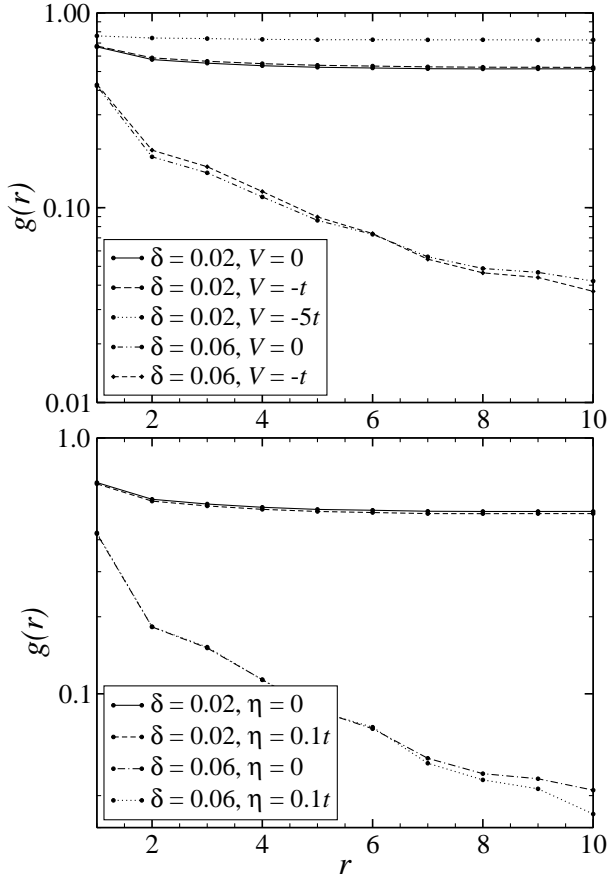}
\caption{ $g(r)$ for $J=0.2t$ in the presence of the out--of--plane
oxygen atoms. The upper panel demonstrates the influence of the
diagonal disorder, whereas the lower panel shows effects coming from
the site--dependent exchange interaction. The model parameters
($\delta,V,\eta$) are given in the legend.} \label{fig7}
\end{figure}

In the upper panel of Fig. \ref{fig7} we show
a comparison of the correlation function $g(r)$ calculated
in the presence of the diagonal disorder and without it.
>From this figure one sees that the influence of the diagonal
disorder is almost negligible, at least for a small--to--moderate
values of the potential $V$. For larger values of $V$, the presence
of negatively charged out--of--plane oxygen atoms reduces the hole
mobility resulting in a visible enhancement of the spin--spin
correlation function. It is worthwhile to emphasize that the NDO
constraint becomes very important in the presence of the diagonal
disorder despite the low concentration of holes. In a homogeneous
systems with $\delta \ll 1$ this constraint is less important as the
probability of double hole occupancy is proportional to $\delta^2$.
Since the negatively charged out--of--plane oxygen atoms locally
enhance the hole concentration, neglecting of the NDO may
significantly modify the results \cite{my}.

Now we turn to the influence of the inhomogeneity--induced
enhancement of the exchange interaction.
Following the results of Ref. \cite{my} we assume $J$ to a be site--dependent
quantity:
\begin{equation}
J_{ij}=J\:\left(1+\eta_{ij}\right),
\end{equation}
where
$$
\eta_{ij}=\left\{\begin{tabular}{ll}
$\eta >0$,\ \ & if there is an out--of--plane oxygen atom \\
& above site $i$ or $j$, \\
0,\ \ & otherwise.
\end{tabular}
\right.
$$
In contrast with the superconducting gap \cite{my}, the AF order is
hardly modified by this mechanism. This can be clearly inferred from
the lower panel in Fig. {\ref{fig7}}. The regime for a magnetic
ordering predicted by many calculations in the $t$--$J$ model
extends to much larger dopings than observed in cuprates and this
discrepancy is sometimes attributed to the inhomogeneities, which
are neglected in many theoretical approaches (see the discussion in
Ref. \cite{spanu}). Although, inhomogeneities are expected to
play an important role in high--temperature
superconductors \cite{dagotto}, our results indicate that their
influence in the AF ordering is rather limited. In particular, we
expect that the inhomogeneities introduced by the out--of--plane
oxygen atoms cannot explain the above discrepancy.

\subsection{Finite--size effects}

Since our analysis has been carried out on finite clusters, it is
necessary to check to what extent the results are affected by the
finite size effects.
One of the measures of
the significance of the finite size effects is a sensitivity 
to the boundary conditions. Therefore, we have calculated the correlation functions and
the spectral functions for systems with different boundary conditions and compared
them to those which have been obtained with periodic boundary conditions. Here,
we have used a method known as averaging over boundary conditions (ABC) \cite{abc}.
Each time 
a particular hole jumps out of the cluster, it is mapped back into
the cluster with wave function with a different phase. Then the
results are averaged over these phases; thereby the reciprocal space
is probed in a much greater number of points than in the case of
periodic boundary conditions. In our simulations we have used a
slightly modified version, where the phases were chosen randomly in
each Monte Carlo step. The averaging over the boundary conditions
has been carried out concurrently with averaging over the
ensemble generated in Monte Carlo run \cite{abc1}. This way it does
not require an additional computational effort.

\begin{figure}[h]
\includegraphics[width=8cm]{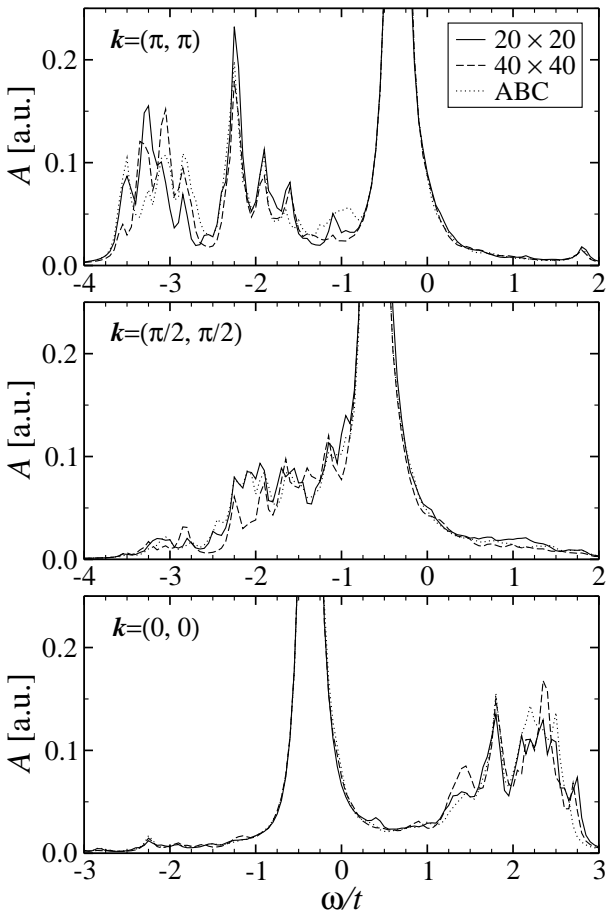}
\caption{Spectral functions calculated for selected points of the
Brillouin zone. These results have been obtained on 20$\times$20
(solid line) and 40$\times$40 (dashed line) clusters with periodic
boundary conditions and on 20$\times$20 cluster with ABC (dotted
line). Since the positions of the coherent peaks obtained with help
of these three approaches are almost indistinguishable, we have cut
the vertical axis in such a way that the incoherent parts are more
pronounced. These results have been obtained for $J=0.2t$,
$kT=0.1t$, and $\delta=0.02$.} \label{fse1}
\end{figure}

\begin{figure}[h]
\includegraphics[width=8cm]{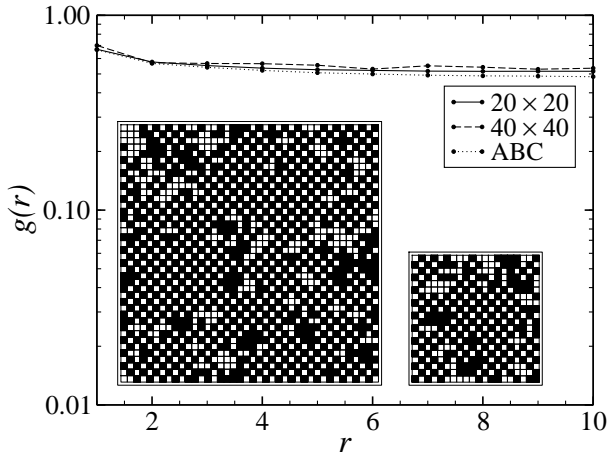}
\caption{Correlation function $g(r)$ for the same systems as in Fig.
\ref{fse1}. Also, the same parameters have been used. The insets
show examples of snapshots of the spin configurations for
40$\times$40 and 20$\times$20 clusters [black (white) square
corresponds to $S_i^z=\frac{1}{2}$ ($S_i^z=-\frac{1}{2})$].}
\label{fse2}
\end{figure}

Another more direct way to check the influence of the finite size
effects is to compare results obtained on clusters of different
sizes. In order to do so, we have repeated some calculations on
40$\times$40 cluster. Fig. \ref{fse1} shows a comparison of spectral
functions obtained for 20$\times$20 and 40$\times$40 clusters with
periodic boundary conditions as well as for 20$\times$20 cluster
with ABC. Since the false--colour plots of these spectral functions
are very similar to each other, we present their energy dependence
for a few selected points of the Brillouin zone.
One can see from this figure that the coherent part of the spectral functions
is almost exactly the same in these three cases. The low--intensity parts
also look very similar, thought some differences can be seen.
Another quantity we have used to analyze the finite size effects is the correlation
function $g(r)$. Fig. \ref{fse2} shows $g(r)$ determined for the same three systems,
for which the spectral functions are presented in Fig. \ref{fse1}.
Despite minor quantitative differences the overall character of all three correlation functions
is the same. Though we have not carried out a systematic finite size scaling, the
similarity of both the spectral functions and the correlation functions constitutes
a significant indication that our results are valid also in the thermodynamic
limit.

\section{Summary}

We have developed a doped-carrier representation of the  Ising
$t$--$J$ model.
In this  formulation, the system  is described in terms
of fermions interacting with static localized spins.
Although it is a slave--particle approach, in contrast with many
similar approaches, the local NDO constraint is taken into account exactly.
The proposed Hamiltonian has the global $Z_2$ symmetry at any values of the parameters, $J$ and $t$.
This model is of an interest in itself since it represents a simple though nontrivial electron system
which captures the physics of strong electron correlations. The issue of how these correlations affect
the magnetic ordering of the lattice spins is thouroughly investigated in the present
work. Besides, this model may provide at least for some values of the model parameters a guess supported
by unbiased numerical calculations regarding the actual low-energy behaviour of the
quasiparticle excitations in a more realistic isotropic $t-J$ model.

In particular, we have calculated the one--hole energy and compared our results with those
obtained for $t$--$J_z$ and $t$--$J$ models. We have found that the one--hole
energy is the same as in the  $t$--$J_z$ model in two limiting cases,
$J\rightarrow 0$ and $J\rightarrow \infty $. For intermediate $J$
the one--hole energy
obtained for the $t$--$J$ model
is in between the results obtained for $t$--$J_z$ and $t-J|_{Z_2}$ approaches and
the differences between $t$--$J$ and $t$--$J_z$ models are comparable
to those between $t$--$J$ and $t-J|_{Z_2}$ ones.

The main advantage of the present approach consists in that it
can be applied for very large systems and the computational effort
increases much slower with the size of the system than, e.g., for exact
diagonalization. Moreover, it works for arbitrary value of the
coupling $J$ and for arbitrary doping level. In particular, in the small--$J$
regime the exact diagonalization and Quantum Monte Carlo methods give rather
poor results. This is because of the fact that in this regime the size of the
defects generated by moving holes is comparable to or larger than the size of
cluster the calculations can be carried on. Since the clusters in our
approach are much larger, this problem is less significant.
Additionally, our method does not require translational invariance
of the system.
This feature is especially important in the context of the recent
experimental results, which clearly indicate the presence of
inhomogeneities  in curates. It could also be applicable to
optical lattices, where the translational symmetry is broken by a
trap.

Using the proposed approach
we have found that the AF SRO persists for
temperatures and dopings which are much beyond the boundaries of the
AF LRO phase,
what is in agreement with recent experiments on the high--temperature
superconductors.
We have also demonstrated that the AF LRO depends on the exchange
interaction $J$. It concerns the transition temperature as well as
the maximal doping at which the AF LRO vanishes.
We explicitly demonstrate that the local no double occupancy constraint
plays the dominant role in destroying the magnetic order at finite doping.

Finally, we have shown that the inhomogeneities induced by
the out--of--plane oxygen atoms have a rather limited influence on
the spin--spin correlations functions, at least in the underdoped
regime. Although, localization of holes by their electrostatic
potential stabilizes the AF LRO, this mechanism becomes important
only for a relatively strong diagonal disorder. Such a limited
influence of inhomogeneities on the AF order is closely related with
the NDO constraint. Note that exactly at half filling the diagonal
disorder does not influence the system, provided the NDO is properly
taken into account.

\ack

This work has been supported by Bogolyubov--Infeld program and by
the Polish Ministry of Education and Science under Grant No. 1~P03B~071~30.


\end{document}